# A novel all-dielectric optical micromixing device for lab-on-a-chip platforms

Adrià Canós Valero[1], Denis Kislov[1], Dmitrii Redka[2] and Alexander S. Shalin[1]

[1]ITMO University, Kronverkskiy prospect 49, 197101, St. Petersburg, Russia

[2]Department of Photonics, Saint-Petersburg State Electrotechnical University

Corresponding author: adria.canos@optomech.ifmo.ru

**Abstract.** The exciting properties of high index dielectric nanoparticles exhibiting both electric and magnetic Mie resonances are nowadays paving the way towards efficient light manipulation at the nanoscale. A commonly disregarded peculiarity of light scattering by Mie particles is their ability to extract angular momentum from the incident electromagnetic field. In this work, we have investigated numerically and analytically the angular momentum transfer from a circularly polarized plane wave to the transversely scattered field generated by a high index cube. Therefore, the tangential components of the Poynting vector enable predominant scattering forces in the near-field and induce orbital motion of absorbing nanoparticles in the vicinity of the scatterer. We then illustrate how these scattering forces can be utilized to realize a simple all-dielectric micromixing scheme for small dipolar spherical Au nanoparticles in an aqueous medium, accounting for the presence of Brownian and drag forces. The novel method we propose represents a step forward towards the practical implementation of efficient, all-dielectric, moving-part free, active mixing devices for a variety of microfluidics applications such as, e.g., lab-on-a-chip platforms for fast chemical synthesis and analysis, preparation of emulsions, or generation of chemical gradients.

## 1. Introduction

Microfluidics is one of the most promising trends in current state-of-the-art science and engineering [1]. In particular, the control of fluid flows in microsized channels plays an essential role for applications ranging from the transport of reduced amounts of hazardous or costly substances and DNA biochip technology, to miniaturized analytical and synthetic chemistry [2–4].

The multidisciplinary nature of microfluidics has allowed to bring together seemingly unrelated fields [1], such as electrical and mechanical engineering, biology or chemistry. For example, in the context of chemical engineering, the utilization of distributed microreactors working in parallel can enhance production significantly and facilitates the design of new products [5,6]. However, slow mixing processes constitute a bottleneck that restricts reaction processes, especially when the desired reaction rate is high [7,8]. For this purpose, fast mixing is highly required to avoid the reactive process being delayed by this critical step, and to reduce potential side products [8].

Given the low Reynolds numbers at which fluid flow occurs in microreactors, fluid mixing represents a significant challenge [1,9,10]. In the most conventional situation where only passive mixing happens, the main driving mechanism corresponds to diffusion (Brownian motion) [1], implying mixing to be at a very low rate. Consequently, the effective distance that the molecules of a fluid need to travel in a mixer before interacting with another fluid with different composition (i.e. - the mixing length), becomes restrictively long [10]. Passive mixers depend solely on decreasing the mixing length by optimizing the flow channel geometry in order to facilitate diffusion [1,11]. In contrast, active schemes rely on external sources injecting energy into the flow in order to accelerate mixing and diffusion processes and drastically decrease the mixing lengths [9,12].

Most early studies related to micromixers have been focused on the passive type. Conversely, despite their higher cost and complex fabrication methods, the enhanced efficiency of active micromixers with respect to passive ones has drawn the attention of the scientific community in the recent years [9]. Because of the power and size constraints involved in microfluidics, research efforts have been focused on the utilization of mixing principles not involving moving mechanical parts such as surface tension-driven flows [13], ultrasound and acoustically induced vibrations [14,15], and electro- and magneto-hydrodynamic action [10,16].

Given the small operation scales of microfluidics, induced optical forces could be particularly relevant, e.g. for micro- and nanoparticles. Consequently, optical manipulation might prove itself as an exciting alternative to the currently investigated active micromixing schemes. The fundamental concept on which the latter could be built-upon is the mechanical momentum transfer from electromagnetic radiation to a structure via scattering and absorption.

Nowadays, a broad range of practical applications rely on the interplay between optics and mechanics [17–20], among them: optical binding schemes [5, 6–9], optical tweezers [22], laser cooling [23], particle sorting [24], and optomechanical light modulation [20]. Consequently, optical manipulation has become a cornerstone to several fields of science and engineering [25]. In order to increase the efficiency and effectiveness of optical manipulation devices, highly localized light profiles with low power trapping beams are currently involved. Complex beam shaping could improve the quality and precision of optical manipulation (e.g., Ref. [26]), however their experimental implementation is much more difficult in comparison with traditional light sources (e.g. linearly and circularly polarized plane waves). Importantly, high light intensities could be harmful for manipulated objects, in particular - for biological specimens [27].

On the other hand, all-dielectric nanophotonics presents itself as a promising alternative for the integration of optomechanics in microfluidic systems. The fields scattered by dielectric nanoparticles with tailored Mie resonances provide the means for tailoring the electric and magnetic components of light [28,29], and can be utilized to induce optical forces on other subwavelength scatterers [30–34]. Due to their inherent low absorption, heat losses to the host environment can be considerably reduced. Relying on the above, the investigation of optical forces induced by carefully engineered high-index dielectrics is of considerable interest and could lead to new remarkable phenomena.

In this work, we study the conversion of Spin Angular Momentum (SAM) of an incident circularly polarized plane wave into Orbital Angular Momentum (OAM) [35,36] via an isolated high-index specially designed nanoparticle. This effect implies the spiral-like profile of the scattered Poynting vector in the transverse plane (an optical vortex). Small absorbing particles placed in the vicinity interact with the scattered field and experience a non-conservative optical torque revolving them around the optimized scattering source. Finally, we propose a novel type of optical active micromixer mediated by chemically inert Au nanoparticles with finely tuned polarizability driving the predominating contribution of rotating force in the near-field (see figure 1). These nanoparticles are involved as intermediate agents for transferring OAM to a fluid and possible admixtures. We believe this work providing a rather simple, optically driven mixing scheme is of high importance for a plethora of applications including microfluidics, lab-on-chip devices, micro- and nanomanipulation etc.

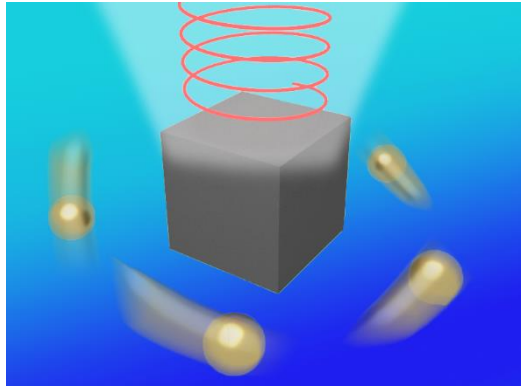

**Figure 1.** Artistic view of the proposed active micromixing scheme. A silicon nanocube submerged in a water solution is illuminated by a circularly polarized light beam. The scattered Poynting vector induces the spiral motion of absorbing Au nanoparticles in the vicinity of the scatterer. Simultaneously, transfer of mechanical OAM takes place between the Au nanoparticles and the fluid, resulting in enhanced convective diffusion and improved mixing.

## 2. Results and discussion

We consider an experimentally realizable geometry known to provide transverse scattering [37] and optimize it for the aforementioned problem taking into account that we are not restricted to the total suppression of forward and backward scattering as in [37], and, therefore, we can tune the excitation frequency in order to obtain a well-defined optical vortex in the transverse plane. Hereinafter, a cubic Si nanoparticle with refractive index $n \approx 4$ (e.g., silicon in the visible range) and 250 nm edge is illuminated by a circularly polarized plane wave propagating along the *z*-axis. For such an incident field one can calculate the angular momentum flux density using the expressions for paraxial waves [38]. The *z* component can then be written in the general form

$$J_z(\mathbf{r}) = \frac{c\varepsilon_0}{2i\omega} \mathbf{E}_{inc}^{*}(\hat{r}\times\nabla)\mathbf{E}_{inc}\bigg|_z + \frac{c\varepsilon_0}{2i\omega} \mathbf{E}_{inc}^{*} \times \mathbf{E}_{inc}\bigg|_z \quad (1)$$

where $\mathbf{E}_{inc}$ is the electric field. The first term in the right-hand side of (1) corresponds to the OAM carried by $\mathbf{E}_{inc}$, while the second term can be interpreted as the SAM flux density. Substituting into (1) the expression of a circularly polarized plane wave yields

$$J_z = \sigma \frac{I_0}{\omega} \quad (2)$$

the OAM term is zero in this case, and the total angular momentum flux density is entirely given by the SAM flux density. The wave helicity $\sigma$ takes the values of +1 and -1 for left and right-circular polarization respectively and $I_0$ is the incident light intensity. Because the total angular momentum is conserved in the scattering of a particle with negligible losses, when the incident field is scattered by the cube, the total angular momentum conservation law implies that part of the SAM is transferred to the orbital and spin terms of the scattered field.

We can write the total angular momentum surface density of the scattered wave in a full analogy with classical mechanics [39] as

$$\langle \mathbf{J} \rangle = \frac{\mathbf{r}\times\langle\mathbf{S}\rangle}{c} \quad (3)$$

$\langle \bullet \rangle$ denotes time average. Since $\mathbf{J}$ is non-zero due to the above, Eq. (3) implies that the Poynting vector of the scattered field has non-zero tangential components.

Following Shamkhi et al.[37], we first focus our attention on spectral points nearly the so-called transverse scattering condition. In our case, however, we consider circularly polarized light. Therefore, the transverse components of the Poynting vector present a remarkable spiral-like behavior. This rotating Poynting vector pattern displays similar features to the ones found in chiral objects such as helices or gammadion-like structures [40–42], but unlike them the fabrication process for our cubic particles is much less complex. The nature of the effect necessarily holds a relation with the multipolar content of the scattered field. In order to gain better physical insight, we consider the multipole decomposition for the scattering cross section of the nanoparticle illuminated with LCP light (figure 2(a)) [43]. The consequent numerical optimization shows the maximal enhancement to take place at a slightly shifted frequency coinciding with the magnetic quadrupole (MQ) resonance (figure 2(b)).

In particular, we emphasize that this magnetic quadrupole mode presents a very high signal-to-noise-ratio with respect to the other leading multipoles; the magnetic dipole and electric quadrupole are almost one order of magnitude smaller and are both out of resonance, while the electric dipole radiation is suppressed by an anapole state coinciding with the magnetic quadrupole resonant frequency. Thus, "pure" magnetic quadrupole fields can be obtained providing stronger near-field effects in comparison with the lower quality resonances [44]. Therefore, the observed spiral-like behavior of the Poynting vector is intrinsically related to the properties of the resonant MQ mode under circularly polarized illumination.

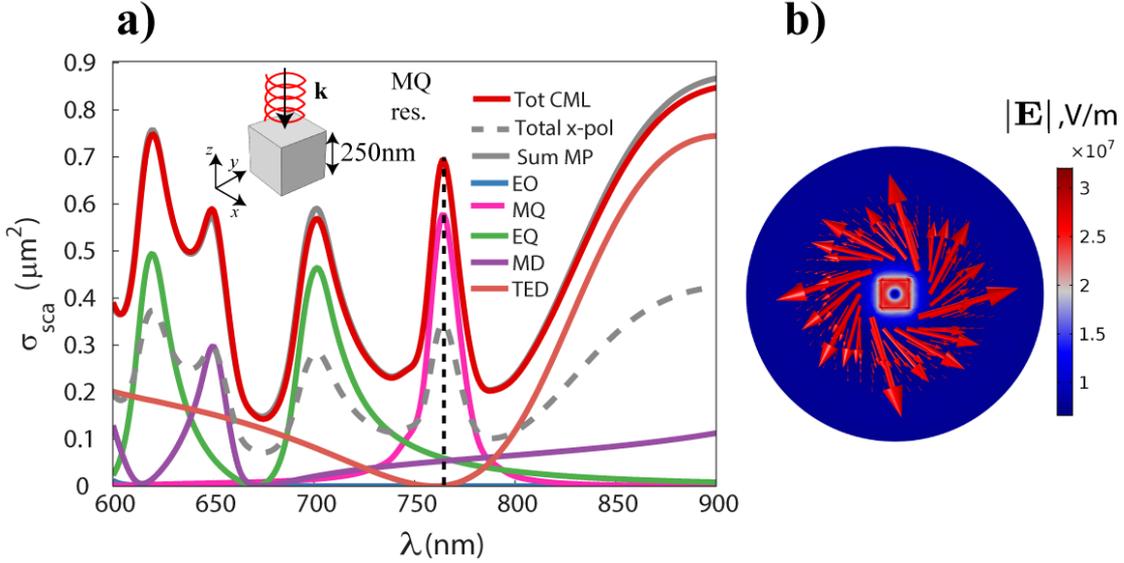

**Figure 2.** (a): The Cartesian multipole decomposition of the dielectric cube scattering cross section in free space under LCP illumination. The dashed black line indicates the position of the resonant MQ mode, and the dashed grey line corresponds to the total scattering cross section when the incident wave is linearly polarized; the total scattering cross section calculated with Comsol (tot CML) is in perfect agreement with the sum of the multipole contributions (sum MP) indicating that the amount of multipoles considered suffices in order to describe the electromagnetic response of the cube. Inset of (a): nanoparticle and incident illumination setup. (b) Colorplot: norm of the total electric field at the resonant wavelength of the MQ in the transverse plane. The arrows show the distribution of the transverse part of the Poynting vector.

While the numerical results shown above provide a clear link between the enhanced SAM to OAM transfer to the scattered field and the magnetic quadrupole resonance, an intuitive physical picture requires further knowledge of the behavior of the fields produced by the excited components of the magnetic quadrupole tensor under the prescribed illumination. For that purpose, we now turn our attention towards the amount of power that the nanocube extracts from the field (the extinction power), defined as [39]

$$P_{ext} = \frac{1}{2}\text{Re}\left\{\int_{V_p} \mathbf{E}_{inc}^* \cdot \mathbf{j}(\mathbf{r}) d^3\mathbf{r}\right\} \quad (4)$$

where the integration is carried over the volume of the scattering source $V_p$ and $\mathbf{j}(\mathbf{r})$ is the induced current inside the scatterer. We consider $\mathbf{E}_{inc}$ to be a left-circularly polarized plane wave with the form $\mathbf{E}_{inc}(\mathbf{r}) = E_0 exp(-ik_0 z)\hat{x} + E_0 exp(-ik_0 z + i\pi/2)\hat{y}$. Performing the multipole expansion of $\mathbf{j}(\mathbf{r})$ in (4)[45], and substituting the expression of $\mathbf{E}_{inc}$, we obtain at the magnetic quadrupole resonance

$$P_{ext} = \frac{k_0^2}{4} \text{Re}\left\{M_{zx}E_y^*(0) - M_{zy}E_x^*(0)\right\} \quad (5)$$

where $M_{ij}$ is the $ij$ component of the magnetic quadrupole tensor and the subscripts $x$ and $y$ for $E$ indicate the $x$ and $y$ components of $\mathbf{E}_{inc}$, respectively. We have placed our particle at the origin of the coordinate system, with the $x$, $y$ and $z$ axis oriented perpendicular to the sides of the cube. Under this conditions, the optical response of the particle for plane waves polarized in the $x$ or $y$ directions is identical, which allows to write $|M_{zy}| = |M_{zx}|$, in full agreement with the numerical results presented in figure 2(a). Further replacing the expressions for the components of the electric field yields

$$P_{ext} = -\frac{E_0 k_0^2}{2} \text{Im}\{M_{zx}\} \quad (6)$$

several conclusions can be readily drawn from (5) and (6). Firstly, due to the chosen excitation wave setup, only the $M_{zx}$ and $M_{zy}$ components of the magnetic quadrupole tensor are excited (since we work with the irreducible Cartesian multipole representation, the multipole tensors are symmetric and therefore the $M_{xz}$ and $M_{yz}$ components are also excited). Secondly, due to the symmetry of the system, their respective contributions to the extinction over an oscillation period are equal. Therefore, the total extinction power at the resonance is exactly two times the amount under linear polarization. In our case, since absorption is negligible, the same applies to the scattering cross section, as shown in figure 2(a). The overall picture can be described as following; during an oscillation period, the incident electric field gradually changes its polarization from one Cartesian axis to the other. Consequently, the excited components of the magnetic quadrupole tensor also change accordingly. The scattered electric field presents mixed contributions from the two magnetic quadrupole components. Moreover, the two components are also $\pi/2$ delayed from each other. In analogy with the case of a rotating electric dipole [46], the scattered near-field at the resonance can be obtained as a superposition of the fields generated by $M_{zy}$ and a $-\pi/2$ delayed $M_{zx}$ component with equal amplitudes. Due to the phase delay, the Poynting vector acquires a nonzero tangential component $S_\phi$ [47]:

$$S_\phi = \frac{3|M_{zx}|^2}{16\pi^2 c\mu_0} \frac{\left(9 + 3r^2 k_0^2 + r^4 k_0^4\right)}{k_0 r^7} \cos(\theta)^2 \sin(\theta) \quad (7)$$

where $\theta$ is the polar angle in spherical coordinates, and $r = \sqrt{x^2 + y^2 + z^2}$. The time-averaged angular momentum density component in the $z$-axis $J_z$ determining the rotation in the transverse plane can be readily calculated substituting the scattered Poynting vector in (3). We find that

$$J_z = \pm \left|\frac{r}{c} S_\phi\right| \quad (8)$$

In the chosen coordinate system, the "+" sign corresponds to illumination with LCP light, and the "–" sign - to RCP illumination. Expression (8) provides direct evidence that SAM from the incident wave has been effectively transferred to the scattered field giving rise to the spiral-like profile observed in the simulations. Moreover, since $J_z$ depends on the choice of origin [48] it can be directly correlated with the extrinsic OAM of the scattered field. In the near field, since the angular momentum scales with $r^{-n}$ ($n$ a positive integer), the vorticity of the Poynting vector is very intense close to the particle, but decreases very fast when going away from it, as confirmed in the simulations (figure 2(b)). Moreover, further inspection of (7) and (8) also shows that the tangential component of the Poynting vector as well as the angular momentum scale quadratically with the magnetic quadrupole moment, enhancing vorticity at the resonance.

Thus, we have provided exact analytical expressions describing the interaction of the magnetic quadrupole mode with circularly polarized incident light, and proposed an intuitive physical picture, from which the origin of the transverse component of the Poynting vector can be understood. Due to the other multipole contributions being negligible, we are free to exploit this effect without further consideration of possible interferences with other modes [49].

### 3. Induced Optical Forces on Small Dipolar Particles

We can now proceed to study the effect of the scattered field on small dipolar particles. In this case, the time-averaged optical force $\langle \mathbf{F}_o \rangle$ can be written as [50]

$$\langle \mathbf{F}_o \rangle = \frac{\alpha'}{2} \sum_{i=x,y,z} Re\left\{E_i^* \nabla E_i\right\} + ck_0\alpha''\left(\frac{n_m^2}{c^2}\langle \mathbf{S}\rangle + \nabla \times \langle \mathbf{L}_s \rangle\right) \qquad (9)$$

where $\alpha'$ and $\alpha''$ denote respectively the real and imaginary parts of the particle dipole polarizability, $n_m$ is the refractive index of the host medium and $\varepsilon_0$ is the vacuum permittivity. The first term in the right-hand side of (9) corresponds to conservative (curl-free) forces acting on the particles. The terms in round brackets are non-conservative or scattering forces, hereinafter noted as $\langle \mathbf{F}_{sc} \rangle$. The latter receives contributions from the Poynting vector and the electric field contribution to the SAM flux density $\mathbf{L}_s$ [50], given by $\mathbf{L}_s = \varepsilon_0 \varepsilon_m \mathbf{E} \times \mathbf{E}^* / (4i\omega)$. When the role of $\mathbf{L}_s$ is assumed negligible compared to the contribution of $\mathbf{S}$, combining Eq. (3) and (9) shows that the $z$ component of the optical torque induced on the particles by $\langle \mathbf{F}_{sc} \rangle$ is proportional to the tangential component of the Poynting vector, or, equivalently, to $J_z$:

$$\Gamma_z = \left(\mathbf{r} \times \langle \mathbf{F}_{sc} \rangle\right)_z \propto r\alpha'' \langle S_\phi \rangle \propto J_z \alpha'' \qquad (10)$$

where $\Gamma_z$ is the $z$ component of the optical torque acting on the particles responsible for their rotation in the transverse plane. Eq. (10) clearly illustrates that the amount of orbital torque transmitted to the particles depends on the properties of the scattered field by means of $\langle S_\phi \rangle$ and the optical response of the particle itself by means of $\alpha''$.

### 4. Optically driven mixer in water environment

We now turn our attention towards the potential applicability of the considered effect as a mixing method for microfluidic reactors. In order to illustrate the concept, the high-index cube is placed in a water host medium ($n_m = 1.34$), containing chemically inert, biologically compatible nanoparticles. The dynamics of the latter will be affected by the optical forces arising due to the interaction with the cube's scattered field together with the Brownian and viscous drag forces induced in the fluid.

The goal is to maximize the ratio $\langle \mathbf{F}_{sc} \rangle / \langle \mathbf{F}_o \rangle$ in order to increase the mechanical orbital torque transferred to the metallic particles, which will then rotate around the cube and act as stirrers, greatly enhancing convective diffusion of any hypothetical admixtures present in the system. However, the scattering force in the dipolar approach can only be dominant if the real part of the polarizability is negligible in contrast to the imaginary one (see eq.(9)).

The obvious and most convenient candidates to act as mixing mediators are gold (Au) nanoparticles, since they would not interact with the chemical and/or biological compounds dissolved in the solutions and are utilized in a broad range of microfluidics applications [51,52] and, in particular, regarding diffusion in vortex-like beams [53]. For simplicity, we assume a spherical shape so that their dipole polarizability can be evaluated analytically with the exact Mie theory formulae by the method described in Refs.[54–56]:

$$\alpha = i \frac{6\pi\varepsilon_0 \varepsilon_d}{k_d^3} a_1 \qquad (11)$$

where $k_d$ and $\varepsilon_d$ are respectively the wavenumber and relative permittivity of water and $a_1$ is the first order electric Mie coefficient [57]. The latter depends on the refractive index contrast between the particle and the medium and the dimensionless parameter $k_d R_p$, $R_p$ - the metallic particle radius. Figure 3 shows the results for Au particles of different sizes in the visible range. The calculations were performed for the well-known optical dispersion properties of bulk Au [58] ($\varepsilon_b^{Au}$), taking into account the Drude size corrections due to the limitation of the electron mean free path in small metallic nanoclusters [58]:

$$\varepsilon_p^{Au}(\omega) = \varepsilon_b^{Au}(\omega) + \frac{\omega_{pl}^2}{\omega(\omega + i\gamma_b)} + \frac{\omega_{pl}^2}{\omega(\omega + i\gamma_b')}$$

$$\gamma_b' = \gamma_b + \frac{0.7 v_F}{R_p} \qquad (12)$$

where $\omega_{pl}$, $\gamma_b$ and $v_F$ are the volume plasmon resonant angular frequency, the damping constant from the free electron Drude model and the Fermi velocity, respectively.

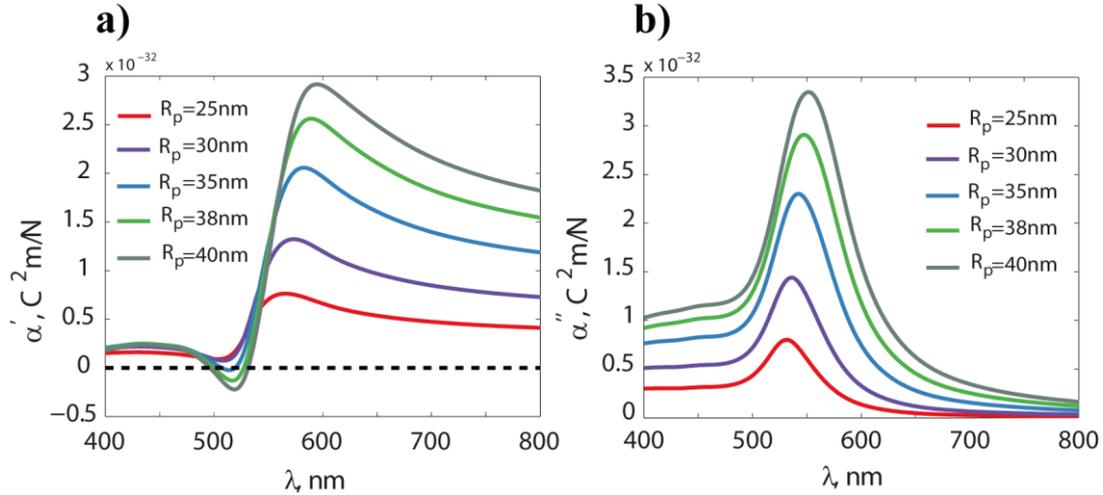

**Figure 3.** Real and imaginary parts of the dipole polarizability for Au nanoparticles with different sizes, calculated with Eqs. (11) and (12). For radius equal or larger than 35 nm zero real part is possible, while the imaginary part of the polarizability is enhanced. The excitation wavelengths correspond to those in free space.

At specific spectral positions in the vicinity of the plasmon resonance, particles with $R_p \geq 35$ nm fulfill the condition $\alpha' = 0$ while presenting enhanced values of $\alpha''$. Consequently, only scattering forces are allowed for these Au nanoparticles. Therefore, for particles with $R_p = 40$ nm, the full suppression of the gradient force occurs at 500 nm and 530 nm (see figure 3(a)). Now, it suffices to scale the size of our cube in order to enable the magnetic quadrupole resonance at, e.g., 530 nm (because of the larger $\alpha''$ in comparison with the 500 nm case).

To predict the trajectories of the Au nanoparticles in the water medium, let us consider the case when the radiation pressure in the direction of propagation is compensated by the substrate presence (the nanoparticles are allowed to move across the outer boundary). Therefore, the problem can be treated as two dimensional in the transverse plane.

As we have already mentioned, we should also take into account Brownian forces in a fluid and viscous drag acting on the nanoparticles. For any particle, we obtain

$$\langle \mathbf{F}_{sc} \rangle + \mathbf{F}_B + \mathbf{F}_D = m_p \ddot{\mathbf{r}}_p \qquad (13)$$

where $\mathbf{F}_B$ is the Brownian force, $\mathbf{F}_D$ is the drag force, $m_p$ is the particle mass and $\ddot{\mathbf{r}}_p$ is the particle instantaneous acceleration vector. Once the scattering force distribution is known, Eq. (13) corresponds to a nonlinear inhomogeneous ODE to be solved in Comsol Multiphysics © utilizing its particle tracing functionality. For small spherical geometries, the latter two forces can be modeled with the expressions [59]

$$\mathbf{F}_B = \phi \sqrt{\frac{12\pi\mu k_B T}{\Delta t} R_p} \qquad (14)$$

$$\mathbf{F}_D = -\frac{18\mu}{4\rho_p R_p^2} m_p \dot{\mathbf{r}}_p \qquad (15)$$

where $\mu$ is the dynamic viscosity of water ($8.9e-4$ Pa·s at ambient temperature), $\phi$ is a dimensionless vector function of randomly distributed numbers with zero mean [59], $T$ is the temperature of the system, $k_B$ is Boltzmann's constant and $\rho_p$ is the density of the particle. The numerical solver models the Brownian forces as a white noise random process with a fixed spectral intensity implying the amplitudes of the force to depend on the iterative time step $\Delta t$ [59]. Formula (15) corresponds to the Stokes Law. Its use is only justified for very low Reynolds numbers in the fluid [60] being, actually, the case for a microfluidic chip [9]. We assume the system to be at ambient temperature.

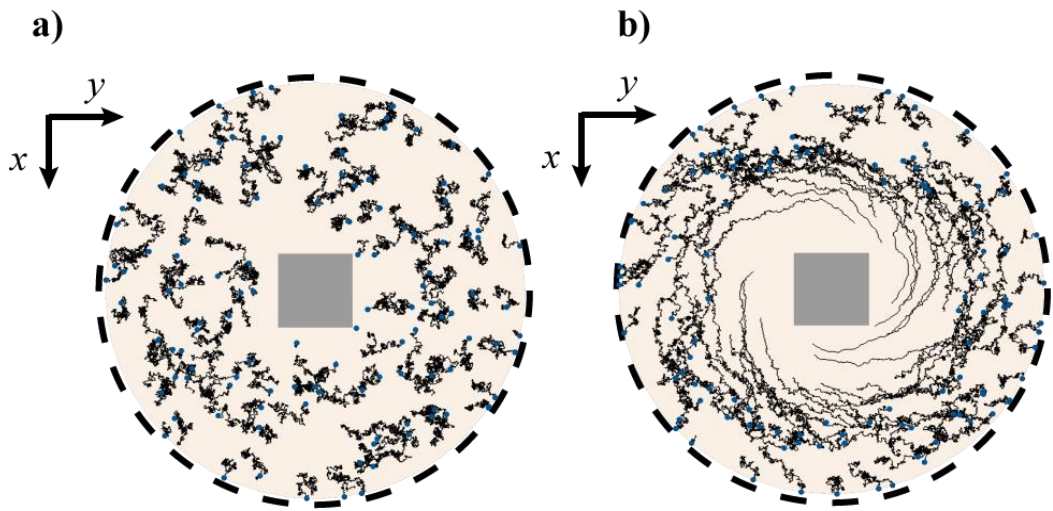

**Figure 4.** Trajectories of Au nanoparticles of 40 nm radius after 0.1ms. Illumination has 530 nm wavelength (value in vacuum), and (a) - only Brownian motion and drag forces act on the particles; (b) - the optical forces are 'switched on', and the Au nanoparticles spirally move outwards the cube. The scaled cube side is 106 nm.

We distribute Au nanoparticles of 40 nm radius uniformly around the Si cube with side 106 nm. Their trajectories after a simulation time of 0.1 ms, are shown in figures 4(a,b). We first consider the situation with Brownian motion and drag forces only (figure 4(a)). As expected, thermal agitation induces random movements of the particles independently on their position in the simulation domain. Conversely, when the system is illuminated with circularly polarized light with an intensity of $50-80$ mW/µm$^2$ (similar to the ones used in conventional optical trapping schemes[61]), a sufficient mechanical torque is transferred to the Au nanoparticles allowing for spiral trajectories (figure 4 (b)).

Eq. (7) reveals $\langle \mathbf{S}_\phi \rangle$ to become negligible in the far field. Therefore, away from the cube the rotating forces are vanishing, and conventional radiation pressure and Brownian motion dominate the dynamics (see figure 5(b)). It is, thus, possible to define an effective radius, $r_m$, limiting the "area of influence" of the optical vortex (red dashed lines in figures 5(a, b)). Figure 5(a) shows the positions of several Au nanoparticles uniformly placed along the *x* axis from the cube. On the one hand, inside the area of influence, the majority of the Au nanoparticles spiral outwards the cube, and the radial distance from the scatterer increases at a slow pace. On the other hand, the behavior of the particles initially placed outside the area of influence is mainly dictated by conventional radiation pressure and thermal agitation (figure 5(b)).

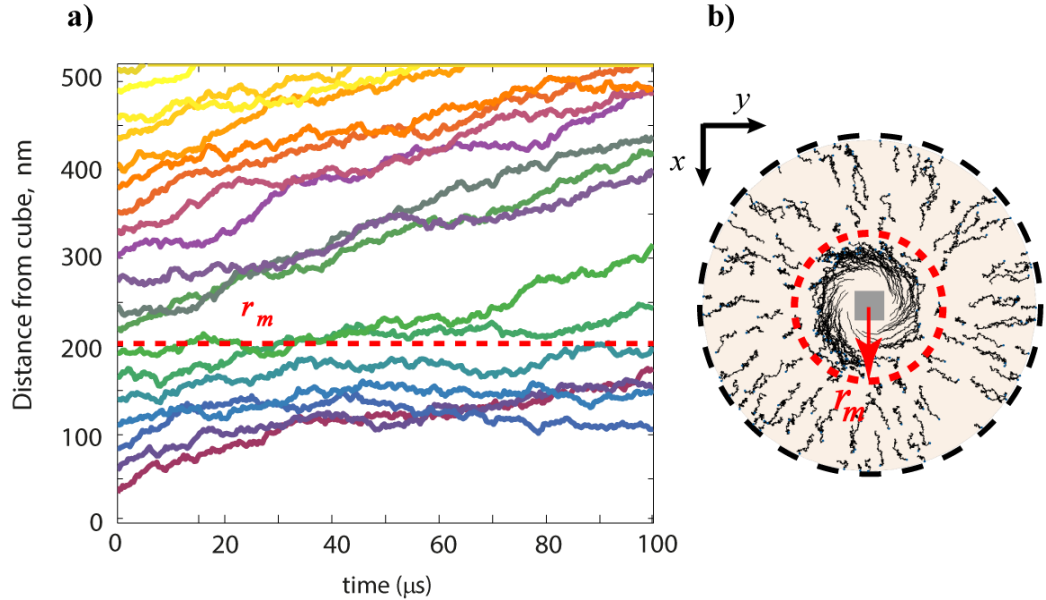

**Figure 5.** (a) Distance from the dielectric cube as a function of time for 19 Au nanoparticles initially evenly distributed along the x axis in the simulation domain depicted in figure 5(b), where the particle trajectories are calculated up to a simulation time of 0.1 ms. $r_m$ is the delimiting radius of the "area of influence" of the individual micromixing element represented by the dashed red line.

Thus, over the range $r_m$ the dielectric cube acts as an effective optical drive for the continuous circulation of the Au nanoparticles which now play the role of stirrers, enhancing convective diffusion in the fluid. In fact, one might also extend our concept for the realization of metasurfaces consisting on periodic arrangements of our dielectric scatterer. Due to the above, our proposed micromixer serves as a fundamental building block for the fabrication of larger micromixing devices allowing the treatment of higher flow input.

Noteworthy that, while all the previous calculations were performed for Au nanoparticles in the visible range, similar dynamics can be also obtained for Ag nanoparticles in the UV range of the spectrum, where they can have zero real part of polarizability [62]. In general, however, UV sources are not always practical because solutions can contain some UV-sensitive substances, such as biological tissues. On the other hand, sometimes Ag-mediated mixing could be extremely useful, e.g., for solutions of photoactivated chemical compounds.

## 5. Conclusion

In conclusion, we obtained the conditions for maximal conversion of SAM of the incident light to OAM of the scattered one via transverse scattering on high-index dielectric cube. The spiral-like profiles of the scattered Poynting vector enabled by the strong magnetic quadrupole resonance drives the rotation of Au nanoparticles in an aqueous solution via curled non-conservative scattering forces. The numerical simulations taking into account both the Brownian and viscous forces demonstrated the effective stirring of the considered admixture in a microchamber. Moreover, the chemically inert metallic particles could mediate the momentum transfer to the fluid boosting the convective diffusion of other components. Therefore, the proposed simple all-optical active mixing scheme does not require moving bulk parts in a microfluidic chamber and could be easily implemented as part of a fabrication process. We believe, the obtained novel phenomena driven by the unique optical signatures of high-index dielectric nanoantennas and peculiar optomechanical properties of small plasmonic nanoparticles could be extremely useful for "lab-on-a-chip" and microfluidics applications. The multidisciplinarity provided by the combination of all-dielectric nanophotonics and optomechanics could pave a way towards enhanced and flexible optically induced operation in closed microchambers for a broad range of applications including speeding-up diffusion, governing chemical reactions, more efficient dynamical sensing, etc.

## Acknowledgements

The authors acknowledge financial support from the Russian Foundation for Basic Research (grants 18-02-00414 and 18-52-00005); the force calculations were partially supported by Russian Science Foundation (Grant No. 18-72-10127).